## 1.1 Beam Dynamics Issues in the SuperKEKB

Demin Zhou, Haruyo Koiso, Akio Morita, Kazuhito Ohmi, Katsunobu Oide, and Hiroshi Sugimoto
KEK, 1-1 Oho, Tsukuba, Ibaraki 305-0801, Japan
Yuan Zhang
IHEP, Beijing, China
Mail to: dmzhou@post.kek.jp

### 1.1.1 Introduction

Assuming 3D asymmetric Gaussian flat beams and neglecting hourglass effect, the luminosity of an electron-positron colliders is given by $\mathcal{L} = \mathcal{L}_0 R_\theta$ with

$$\mathcal{L}_0 = \frac{N_+ N_- f_c}{2\pi \sqrt{\sigma_{x+}^2 + \sigma_{x-}^2} \sqrt{\sigma_{y+}^2 + \sigma_{y-}^2}} \tag{1}$$

and the geometrical reduction factor

$$R_\theta = \frac{1}{\sqrt{1 + \frac{\sigma_{z+}^2 + \sigma_{z-}^2}{\sigma_{x+}^2 + \sigma_{x-}^2} \tan^2\left(\frac{\theta}{2}\right)}}, \tag{2}$$

where the subscripts of $+$ and $-$ respectively denote the positron and electron bunches, $f_c$ is the bunch collision frequency, $N$ is the bunch population, $\sigma$ is the beam size in the horizontal ($x$), vertical ($y$) and longitudinal ($z$) directions, and $\theta$ is the full crossing angle. The transverse beam sizes at the interaction point (IP) are determined by $\sigma_{x,y}^* = \sqrt{\beta_{x,y}^* \epsilon_{x,y}}$, where $\epsilon$ is the beam emittance, and $\beta$ is the beta function at the IP in each plane.

With the same assumptions for the luminosity formula, the beam-beam tune shifts for the electron beam are expressed by

$$\xi_{x-} = \frac{N_+ r_e \beta_{x-}}{2\pi \gamma_- \bar{\sigma}_{x+} (\sigma_{y+} + \bar{\sigma}_{x+})}, \tag{3}$$

$$\xi_{y-} = \frac{N_+ r_e \beta_{y-}}{2\pi \gamma_- \sigma_{y+} (\sigma_{y+} + \bar{\sigma}_{x+})}, \tag{4}$$

where $r_e$ is the classical radius of electron, $\gamma_-$ is the Lorentz factor for the electron beam, and $\bar{\sigma}_{x+} = \sigma_{x+}\sqrt{1 + \Phi_+^2}$ with Piwinski angle $\Phi_+ = \frac{\sigma_{z+}}{\sigma_{x+}} \tan\left(\frac{\theta}{2}\right)$. Reversing the signs of $+$ and $-$, the above equations also hold for the positron beam. For particles with zero longitudinal displacements in the electron beam, they feel a positron bunch with effective width of $\bar{\sigma}_{x+}$. The beam-beam tune shifts $\xi_{x,y}$ denote the strength of the interaction of colliding beams, and usually saturate at finite values while beam currents increase and transverse beam size blowup happens. With flat beams, usually the luminosity is mostly sensitive to the vertical beam sizes. For SuperKEKB, with large crossing angle and small vertical beta function, the beam-beam tune shift in the horizontal direction is much smaller than that in the vertical direction, namely $\xi_{x/y+} = 0.003/0.088$ and $\xi_{x/y-} = 0.001/0.081$ for positron and electron beams, respectively.

The SuperKEKB is designed with the strategy of so-called nanobeam scheme, which was originally proposed by P. Raimondi for SuperB [1]. The electron and positron beams collide with a horizontal crossing angle of $\theta = 83$ mrad. The horizontal emittances are $\epsilon_{x+} = 3.2$ nm and $\epsilon_{x-} = 4.6$ nm, taking into account the intra-beam scattering effects. The beam sizes at the IP are $\sigma_{x+} = 10.1$ $\mu$m and $\sigma_{x-} = 10.7$ $\mu$m. The overlap area of the two beams is $\Delta s = \frac{2\sigma_x}{\theta} \approx 0.25$ mm, which is about 1/20 of the nominal bunch length. Another feature of SuperKEKB is the small vertical beta function at IP, which is squeezed to be $\beta_{y+}^* = 0.27$ mm and $\beta_{y-}^* = 0.3$ mm, comparable to the overlap area $\Delta s$. The vertical emittances are assumed to be $\epsilon_{y+} = 8.64$ pm and $\epsilon_{y-} = 12.9$ pm, taking into account various intensity-dependent and -independent effects. Comparing with its predecessor, the emittances of SuperKEKB rings are about 1/5 and 1/20, and the beta functions are about 1/40 and 1/50 of those of KEKB rings in the horizontal and vertical directions, respectively.

Since the KEKB rings, as reviewed in Refs. [2-4], have experienced many beam dynamics issues which affected the luminosity performance, it is expected that the luminosity performance of SuperKEKB will be even more sensitive to various imperfections or perturbations, such as machine errors, lattice nonlinearity, intra-beam scattering, beam-beam interaction, space charge, impedance-driven instabilities, etc. Regarding to the beam dynamics issues associated to the electron-positron colliders, there are reviews in Refs. [5-7]. The progress of next generation B-factory projects have been reviewed in Refs. [8-10], and in Ref. [11] especially the most recent status of SuperKEKB. This paper is dedicated to discussing a few beam dynamics issues which might set challenges for the future commissioning of the SuperKEKB. For more information of beam dynamics issues at SuperKEKB, such as intra-beam scattering, electron cloud effects, impedance effects, optics optimization, etc., the interested readers are directed to Refs. [12-16].

### 1.1.2 Beam Dynamics Issues

This section gives a brief overview of some important beam dynamics issues in SuperKEKB. Since most of these issues appear to be more prominent in the low energy ring (LER), we mainly use the LER for illustrations rather than the high energy ring (HER) in the following discussions.

#### 1.1.2.1 *Intra-beam scattering*

In the case of a low emittance ring with high bunch current, the effect of intra-beam scattering becomes significant, and can affect the equilibrium emittance, bunch length and energy spread. Figures 1 and 2 show these parameters as functions of the bunch population while keeping the nominal ratio of the vertical emittance fixed for SuperKEKB rings. The situations show similar emittance growth as that in SuperB [17]. Especially, in the LER there is an increase of more than 60% in the horizontal emittance, and about 5% in the bunch length and energy spread at the nominal bunch population of $9.04 \times 10^{10}$.

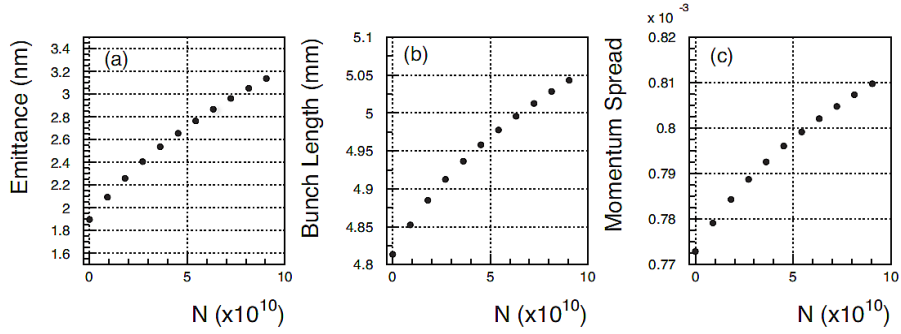

**Figure 1:** Effects of intra-beam scattering in the LER: (a) emittance, (b) bunch length, and (c) energy spread as a function of bunch population.

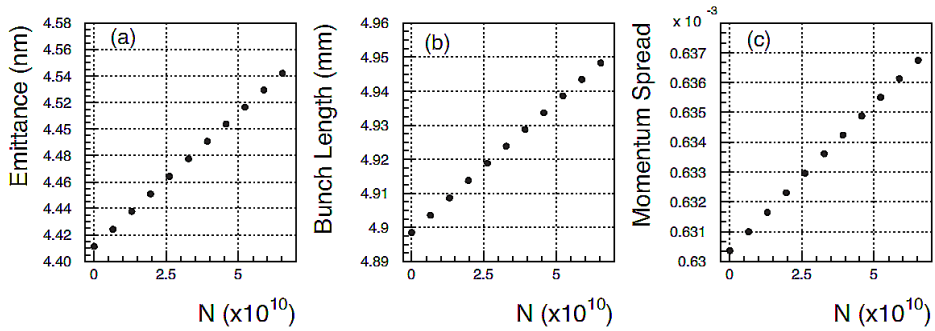

**Figure 2:** Effects of intra-beam scattering in the HER: (a) emittance, (b) bunch length, and (c) energy spread as a function of bunch population.

#### 1.1.2.2 *Beam-beam interaction*

It is well accepted that the 'sweet spot' in the tune space for achieving highest luminosity at an electron-positron collider usually locates at an area close to half integer. To search for the best working point in the tune space, luminosity scans are performed for both LER and HER, with the fractional tunes in the range of [0.5, 0.75] and the beam currents set to design values. The tune scan results of luminosity using a weak-strong model for the LER and HER are demonstrated in Fig. 3 with scaled colours, and Fig. 4 shows the relevant scans of vertical rms beam sizes. It is seen that the strong synchro-betatron resonances of $2\nu_x - N\nu_s = Integer$ exist in the nanobeam scheme. This is due to the large crossing angle chosen for the purpose of mitigating hourglass effects. Furthermore, the resonances of $\nu_x + 2\nu_y + N\nu_s = Intege$, $2\nu_y - N\nu_s = Integer$, and $\nu_x - \nu_y - \nu_s = Intege$ also restrict the choice of working point. The working points have to be kept far enough from these strong resonances. In general, the luminosity is very sensitive to the vertical beam size. With higher beam energy, the electron beam in HER is more robust than the positron beam in LER with respect to the beam-beam driven synchro-betatron resonances. At present, both the main rings of the SuperKEKB are optimized with fractional tunes of [0.53, 0.57]. The working point is selected from islands isolated by the beam-beam resonance lines. But notice that the island areas might shrink when the lattice nonlinearity and machine errors strengthen those resonances [13].

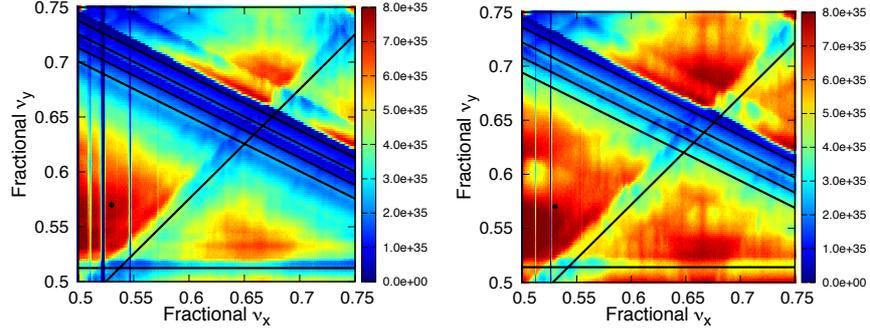

**Figure 3:** Tune scan of luminosity for the LER (left) and HER (right). The black dots and lines indicate the nominal working point and the various synchro-betatron resonances.

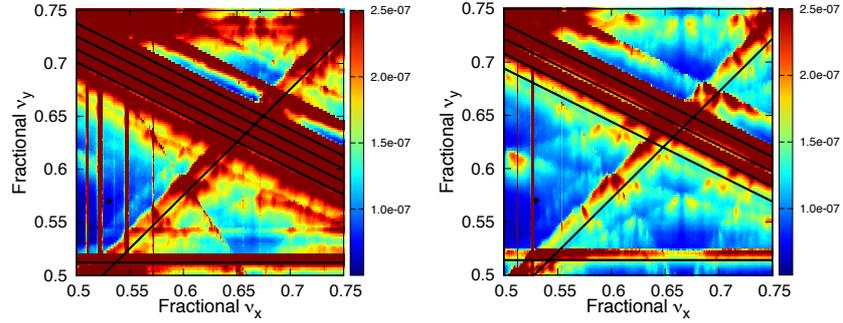

**Figure 4:** Tune scan of vertical rms beam size at IP for the LER (left) and HER (right). The black dots and lines indicate the nominal working point and the various synchro-betatron resonances.

#### 1.1.2.3 *Lattice nonlinearity*

For SuperKEKB, most of the unavoidable lattice nonlinearity is attributed to the interaction region resulting from the extremely small beta functions at IP and low emittances. For examples, the nonlinear terms from the drift space near IP, so called kinematic terms, and the Maxwellian fringe fields of final focus (FF) superconducting quadrupoles will become very important. The dynamic aperture (DA) limited by these terms in a circular collider has been studied in Refs. [18, 19]. The aperture in term of initial action variable is written as

$$J_y = \frac{\beta_y^{*2}}{(1+2|K_1|L^{*2}/3)L^*} A(\mu_y),  \quad (5)$$

where $L^*$ is the distance from the IP to the quadrupole face, $K_1$ is the quadrupole strength, and $A(\mu_y)$ is a universal function in term of vertical phase advance $\mu_y$, which has a meaning of tune shift in Ref. [19]. The relevant parameters are summarised for some colliders in Table 1, where the value of the scaling factor $J_y/A(\mu_y)$ indicates the difficulty of achieving large DA. It turns out that the SuperKEKB has the smallest $\beta_y^*$, and is likely the most challenging project. It is also noteworthy that, different from other projects, the FF quadrupole fringes of SuperKEKB are very strong and its effect on DA is even more severe than the kinematic terms.

**Table 1:** Important parameters limiting the dynamic aperture for some colliders. The parameters for CEPC and TLEP are typical design values.

| Ring | $\beta_y^*$ [mm] | $K_1$ [m$^2$] | $L^*$ [m] | $J_y/A$ [μm] |
|---|---|---|---|---|
| SuperKEKB HER | 0.3 | -3.1 | 1.22 | 0.018 |
| SuperKEKB LER | 0.27 | -5.1 | 0.76 | 0.032 |
| CEPC | 1.2 | -0.176 | 1.5 | 0.76 |
| TLEP | 1 | -0.16 | 0.7 | 1.36 |
| KEKB | 5.9 | -1.779 | 1.762 | 4.22 |

In addition to the kinematic and FF quadrupole fringes, there are other important sources of lattice nonlinearity resulting from solenoids of 1.5 T field installed for particle detection. And anti-solenoid magnets, which almost overlay with FF quadrupoles, are adopted to compensate the detector solenoid fields. Due to the large crossing angle, the solenoid axis deviates from the beam axis, that generate unwanted fields acting on the beam. The fringe fields of the solenoids can induce the vertical emittance. The beam orbit is curved due to solenoid field in the LER. Consequently, the FF quadrupoles are shifted downside by 1.5 mm in the left side and 1.0 mm in the right side in order to reduce the dipole angle of the corrector coil to adjust the orbit as small as possible. The rotation of the FF quadrupoles around the beam axis and the skew quadrupole correctors are adopted to make the vertical dispersions and the X-Y couplings in IR as small as possible.

Furthermore, the natural chromaticity in SuperKEKB is very large, and approximately 80% of it in the vertical direction is generated by the FF system. So strong sextupoles are installed for chromaticity correction. To suppress the nonlinearity caused by the FF system, correction coils from dipole to octupole components are installed to each FF quadrupoles [20]. Even so, the IR is not transparent for off-momentum or large-amplitude particles. This is illustrated in Fig. 5. A particle is tracked through the LER with synchrotron radiation excitation and damping turned off. The initial conditions are varied by shifting the initial horizontal coordinates. When we observe the vertical motion in the phase space, we do see the vertical amplitude and even the closed orbit grow while the horizontal amplitude increases (see the middle in Fig. 5). This is a clear evidence of nonlinear horizontal-to-vertical coupling. We do the same tracking for a simplified lattice where we remove the solenoids, and simplify the arrangements on FF quadrupoles. It turns out the X-Y coupling disappears. So, we conclude that the solenoids do contribute to lattice nonlinearity in SuperKEKB.

#### 1.1.2.4 *Space charge*

The first-order space-charge tune shift experienced by particles performing small oscillations around the beam centroid in a uncoupled lattice and for a Gaussian bunch can be estimated by

$$\Delta v_i = -\frac{1}{4\pi} \frac{2r_e}{\beta^2 \gamma^3} \int_0^C \frac{\lambda(s)\beta_i}{\sigma_i(\sigma_x+\sigma_y)} ds, \quad (6)$$

with $\beta_x$, $\beta_y$ the beta functions, $\sigma_x$, $\sigma_y$ the horizontal and vertical rms beam sizes, $C$ the circumference of the ring, $\beta$ the relative velocity, and $i = x, y$. The longitudinal peak density is $\lambda(s) = \frac{N}{\sqrt{2\pi}\sigma_z(s)}$ with gaussian bunch profile assumed. In the absence of linear coupling, the horizontal beam sizes are calculated from emittance via $\sigma_x^2 = \epsilon_x \beta_x + \langle \delta^2 \rangle D^2$ with $D$ the dispersion function.

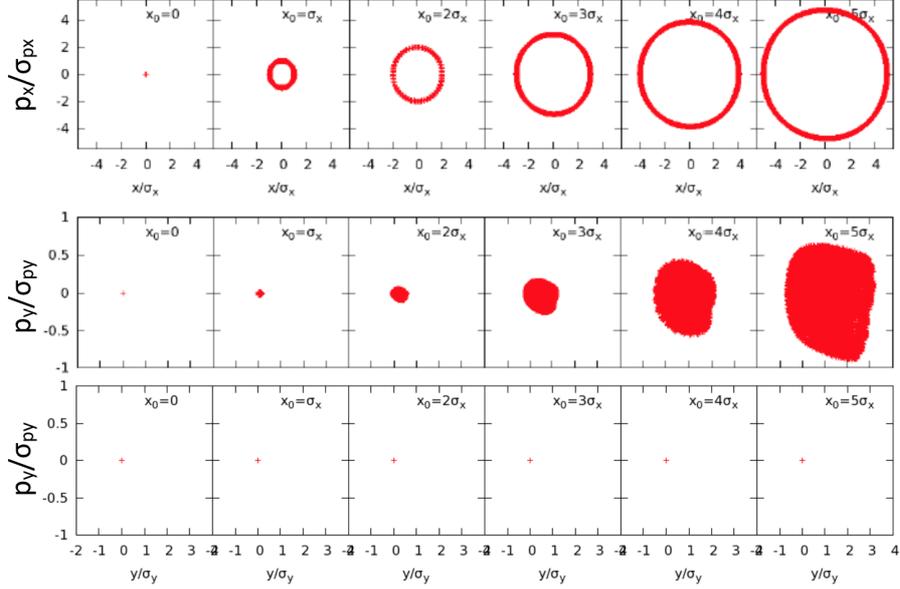

**Figure 5:** Poincare maps at the IP with increasing horizontal offset from the left to the right. Top: horizontal phase space. Middle: vertical phase space for a baseline lattice. Bottom: vertical phase space for a simplified lattice.

Table 2 summarises the estimated beam-beam parameters $\xi_{x,y}$ and linear space charge tune shifts $\Delta \nu_{x,y}$ for SuperKEKB and KEKB rings, compared with the beam-beam tune shifts. It is seen that the space charge tune shifts are very small for KEKB. But for SuperKEKB LER, the space charge tune shift in the vertical direction is in the same level as beam-beam tune shift with opposite sign. Though the linear part can cancel each other, the amplitude-dependent tune shifts will not due to their different nonlinear behaviours. The betatron tune footprints for the LER with and without space charge are shown in Fig. 6. The simulations are done using SAD code [21] based a weak-strong model for space charge effect [22]. With working point set to be close to half-integer for seek of good luminosity, the particles will attracted to half-integer resonance and become unstable.

**Table 2:** Estimated beam-beam parameters and linear space charge tune shifts for the SuperKEKB and KEKB rings.

| Parameter | SuperKEKB | | KEKB | |
|---|---|---|---|---|
| | LER | HER | LER | HER |
| $\epsilon_x$ (nm) | 3.2 | 4.6 | 18 | 24 |
| $\epsilon_y$ (pm) | 8.64 | 11.5 | 180 | 240 |
| $\xi_x$ | 0.0028 | 0.0012 | 0.127 | 0.102 |
| $\xi_y$ | 0.088 | 0.081 | 0.129 | 0.09 |
| $\Delta \nu_x$ | -0.0027 | -4e-4 | -5e-4 | -3e-5 |
| $\Delta \nu_y$ | -0.094 | -0.012 | -0.0072 | -4e-4 |

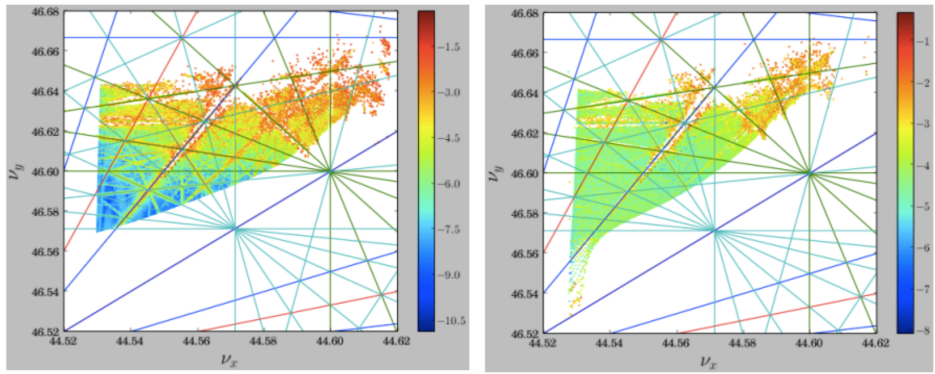

**Figure 6:** Betatron tune footprint for a baseline lattice of LER without (left) and with (right) space charge effect. Resonance lines from 4th to 7th orders are also plotted.

### 1.1.3 Interplay of Beam-Beam With Lattice Nonlinearity and Space Charge

#### 1.1.3.1 *Lattice nonlinearity*

The interplay of beam-bam and lattice nonlinearity in SuperKEKB has been studied in Ref. [13], and in other machines it is reviewed in Ref. [23]. We found that their interplay can spread the betatron tune footprint over an extended area, and cause remarkable luminosity loss at SuperKEKB. For the LER, the main source was attributed to the amplitude-dependent nonlinearity. Here we present more results of its effects on DA, lifetime and beam tails.

Figures 7 and 8 show the DA calculated by particle tracking using SAD for the LER and HER with and without beam-beam interaction [16]. For the LER, The DA is reduced significantly compared with that without the beam-beam effect. The Touschek lifetime is calculated from the fitted sizes of DA. Without beam-beam, the Touschek lifetime is about 600 s for both LER and HER, and almost satisfies the requirement from injection. But with beam-beam, the Touschek lifetime will reduce by around 10% and 85% for HER and LER, respectively. Moving the working point to be closer to half-integer can recover the lifetime for LER, but the loss rate is still more than 50%.

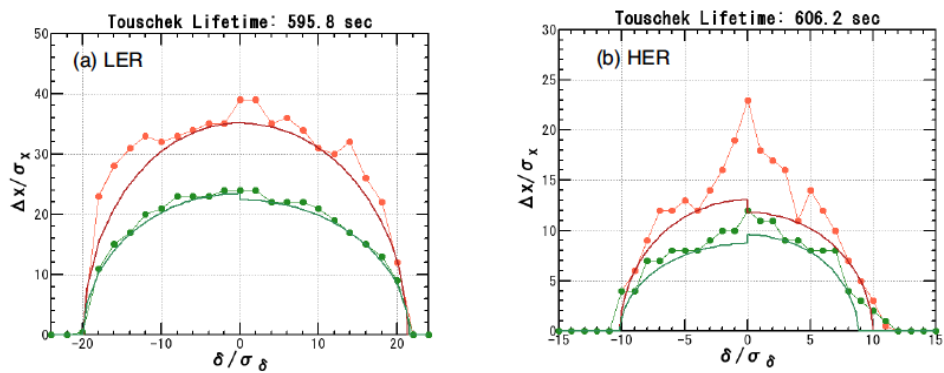

**Figure 7:** Dynamic aperture for LER (left) and HER (right) without beam-beam.

Beam-beam interaction with large crossing angle can generate beam tails [24] that is unwanted by the particle detector. From the comparison of beam tail simulations with pure beam-beam interaction and with lattice nonlinearity included, see the left figures of Figs. 9 and 10, the interplay of beam-beam and lattice nonlinearity does enhance the beam tail of the positron beam in LER. This will cause additional challenges to the collimation system for protection of particle detector.

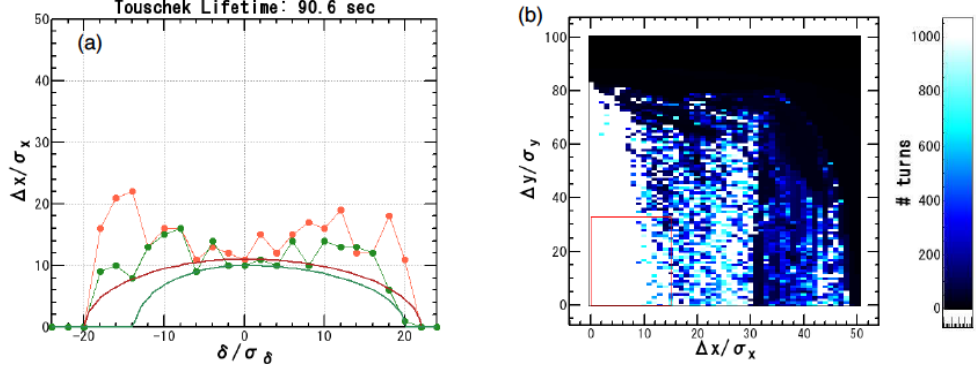

**Figure 8:** Dynamic aperture for LER with beam-beam in the horizontal-momentum plane (left) and in the horizontal-vertical plane (right). The red square indicate the required injection aperture.

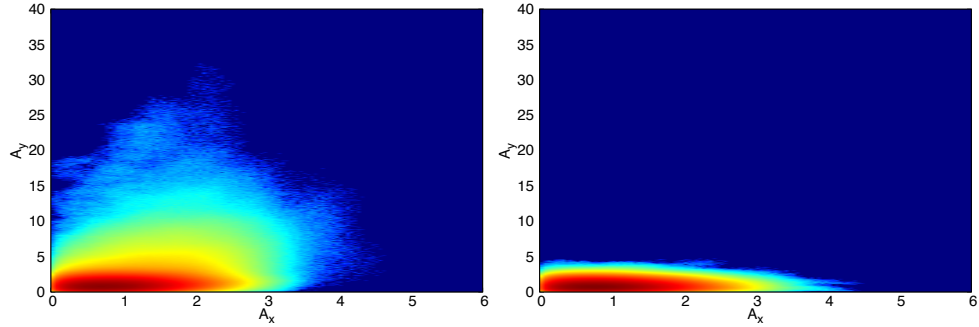

**Figure 9:** Beam tail simulated by BBWS without (left) and with (right) crab waist.

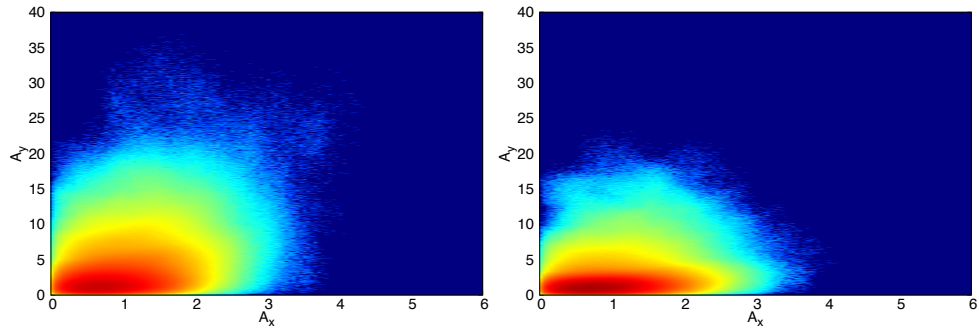

**Figure 10:** Beam tail simulated by SAD for a baseline lattice of LER without (left) and with (right) ideal crab waist.

### 1.1.3.2 *Space charge*

As shown in Fig. 6, the space charge forces drive the particles toward half-integer while beam-beam acts on the opposite direction. When these two forces add on each other, they will create a strongly distorted footprint for particles in the tune space, see Fig. 11. Note that here we used a weak-strong model for space charge, and the blow up in beam sizes is not taken into account. Therefore the simulations are not self-consistent, and overestimate the space charge effects.

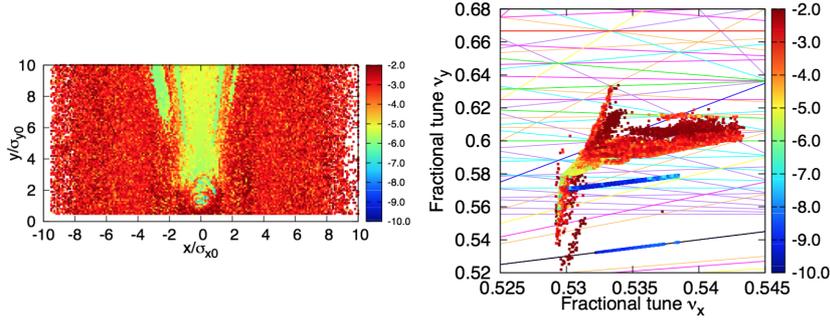

**Figure 11:** FMA for a baseline lattice with space charge. Left is for the physical space and right is for the tune space. Resonance lines up to 10th orders are also plotted.

The luminosity performance as a function of bunch current products are simulated using BBWS and SAD, as shown in Fig. 12 and also in Ref. [13]. Adding to the lattice nonlinearity, space charge does cause additional luminosity loss, though its effect is overestimated due to the use of a weak-strong model. It is interesting to observe that SC does compensate BB effect at low beam currents (see the left of Fig. 12), where the nonlinear effects of these two forces are not significant. For a simplified lattice with solenoids removed and FF quadrupoles simplified, see the right of Fig. 12, the interplay of BB with lattice nonlinearity and SC relaxes in response to less nonlinearity. Therefore, it is concluded that the nonlinear fields from solenoids and consequent re-arrangements of FF quadrupoles play an important role at SuperKEKB.

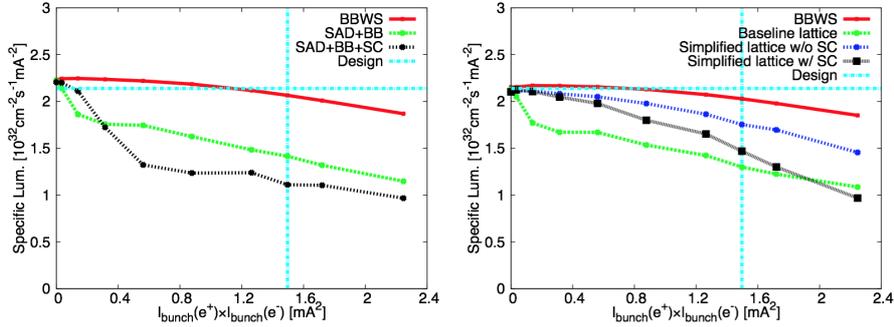

**Figure 12:** Specific luminosity as a function of bunch current products for the LER. The cyan lines indicate design values. The left and right figures have slight differences in nominal beam parameters.

#### 1.1.3.3 *Detuned lattice*

Detuned lattices will be used in the phase 2 operation of SuperKEKB without VXD detector. For these lattices of LER and HER, the vertical and horizontal beta functions at the IP will be 4 and 8 times the values of baseline lattice. We check the effects of beam-beam, lattice nonlinearity and space charge for the detuned lattice of LER, and the results are summarized in Fig. 13. It turns out both space charge and lattice nonlieanrity will be much less important for the detuned lattices. Achieving the target luminosity of $1\times10^{34}$ cm$^{-2}$s$^{-1}$ is very promising, and reaching the value of $1\times10^{35}$ cm$^{-2}$s$^{-1}$ might also be possible by increasing the beam currents.

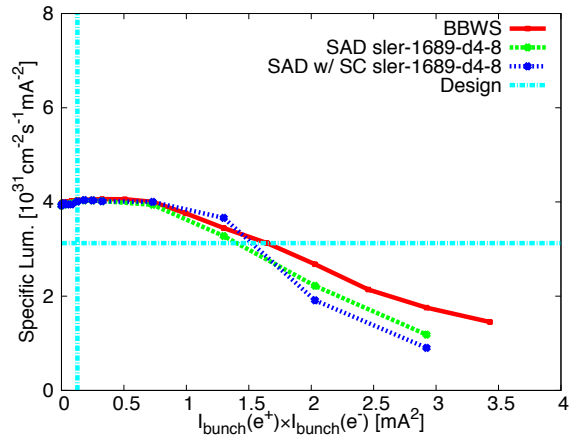

**Figure 13:** Specific luminosity as a function of bunch current products for the detuned lattice of LER. The cyan lines indicate design values.

### 1.1.4 Mitigation Schemes

The crab waist is the most promising technique for suppressing the beam-beam resonances in the nanobeam scheme [1, 5, 24]. As stated in Ref. [5], the crab waist transformation gives a small geometric luminosity gain (around 10% for SuperKEKB) due to the vertical beta function redistribution along the overlap area. However, the dominating effect comes from the suppression of betatron and synchro-betatron resonances arising from the vertical motion modulation by the horizontal betatron oscillations. This is demonstrated by the weak-strong simulations for the positron beam as shown in Figs. 9 and 14. At the same time, there will be more choices for working point in the tune space.

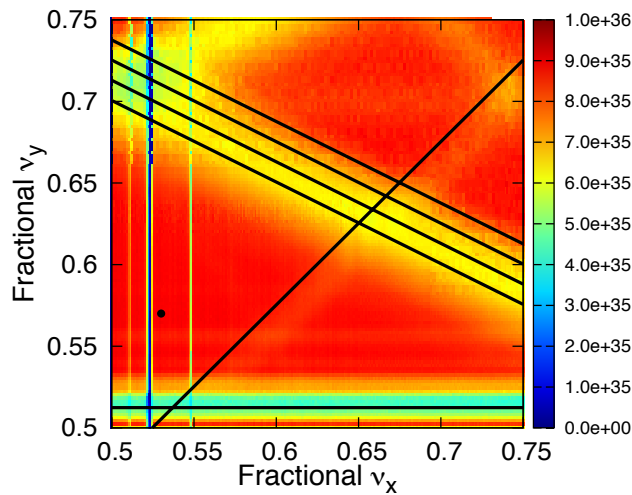

**Figure 14:** Tune scan of luminosity for LER with crab waist. The black lines indicate various resonances.

In the present design of SuperKEKB, crab waist is not adopted because the lattice nonlinearity in the IR is very strong and always cause severe loss of DA and lifetime when crab waist sextupoles are put into the real lattice. For detailed studies of crab waist scheme applied to SuperKEKB, see Ref. [16]. Even with ideal crab waist put at the IP in a real lattice, lattice nonlinearity still can weaken the its power in suppressing in the beam-beam tails, as shown in the right figure of Fig. 10. All studies strongly suggest that the nonlinear optimization of the real lattices is a must for successful application of crab waist to SuperKEKB. Unfortunately, this is not very successful up to now. We expected advanced nonlinear analysis techniques, see Ref. [25] for example, applied to the SuperKEKB lattices. For space charge effects, compensation of the linear tune shift is not enough. The amplitude-dependent tune shift also needs to be compensated by dedicated magnets such as octupoles.

### 1.1.5 Summary and Future Plans

The recent design studies of SuperKEKB show that many beam dynamics issues might affect its final luminosity performance and set challenges to the beam commissioning. For examples, the lattice nonlinearity set limit to the dynamic aperture and Touschek lifetime, interplay with beam-beam and cause luminosity loss, and impede the success of applying crab waist. In the LER, space charge is a new issue and its importance has just been recognized recently.

To remedy these challenges, we plan to perform detailed analysis of lattice nonlinearity in SuperKEKB under an international collaboration program. Connecting the ongoing study efforts on SuperKEKB with the design efforts of future circular colliders will benefit both sides.

### 1.1.6 Acknowledgements

The author D.Z. would like to thank the SuperKEKB team for constant support of this work. Special thanks aredue to M. Zobov, Y. Cai, D. Sagan, and A. Chao for helpful discussions.